\documentstyle{europhys}

\def\stars{\bigskip\centerline{***}\medskip}

\newif\ifboo \boofalse

\def\Review#1{\boofalse{\it #1},}
\def\Name#1{{\sc #1},}
\def\Vol#1{\ifboo Vol. {\bf #1}\else{\bf #1}\fi}
\def\Year#1{\ifboo #1\else(#1)\fi}
\def\Book#1{\bootrue{\it #1},}
\def\Page#1{\ifboo {\rm p. #1}\else{\rm #1}\fi}

\begin{document}
\euro{}{}{}{}
\date{\today}

\title
{
Universal statistics of wave functions in 
chaotic and disordered systems
}

\author
{
Bambi
Hu,\inst{1,2}
Baowen Li,\inst{1,3}\footnote{To whom correspondence should
be addressed. E-mail:
bwli@phibm.hkbu.edu.hk}
and Wenge Wang\inst{1,4}
}
\institute{
\inst{1}Department of Physics and  Centre for Nonlinear Studies, Hong
Kong Baptist University, Hong Kong, China \\
\inst{2} Department of Physics, University of Houston, Houston TX
77204-506, USA\\
\inst{3} Department of Physics, National University of Singapore, 119620 
Singapore\\
\inst{4} Department of Physics, South-east University, Nanjing 
210096, China 
}

\rec{ }{ }
\pacs{
\Pacs{05}{45Mt}{Quantum chaos}
\Pacs{71}{23-k}{Electronic structure of disordered solids}
\Pacs{72}{15Rn}{Localization  effects}
}
\maketitle

\begin{abstract}
We study a new statistics of wave functions in 
several chaotic  and  disordered
systems:  the random matrix model, band 
random matrix model, the Lipkin 
model, chaotic quantum billiard and the 1D tight-binding model. Both 
numerical and analytical results show 
that the distribution function of a 
generalized Riccati variable, defined as
the ratio of components of eigenfunctions on basis states coupled by 
perturbation, is universal, and has the form of Lorentzian distribution. 
\end{abstract}

Statistics of eigenfunctions in disordered and chaotic systems are of 
great interest in many branches of physics such as condensed matter 
physics, nuclear physics, chemical physics, and in particular, in the 
new developing field of quantum 
chaos\cite{WVPRL,Haake,GVZ91}. However, although 
eigenfunctions contain more information 
than eigenenergies, much less has been studied compared with the 
eigenenergies.

The mostly studied property is the statistics of wave function 
amplitude $|\psi({\bf r})|^2$ and two 
point correlation function $\langle \psi({\bf r_1})\psi({\bf r_2})\rangle$. 
The former one has been found to be 
the Porter-Thomas distribution for time-reversal invariant systems whose 
classical counterpart is chaotic. This is
predicted by the Random Matrix Theory \cite{PT,Berry77,Voros79}, and 
confirmed  numerically\cite{Li,Num} and experimentally\cite{Exp}. The 
later one, which is proportional to $J_0(kd)$ ($J_0$ is the Bessel 
function of the zeroth order, $k$ the wave vector and $d$ the distance 
between 
the two points), was proposed by Berry\cite{Berry77} for a chaotic billiard 
within an
assumption that the wave function is a superposition of plane waves with 
random coefficients. The two point correlation function 
has been extended to large separation by Hortikar and 
Srednicki\cite{Srednicki} recently. A beautiful review for the 
correlations of 
wave functions in disordered systems has been given recently by Mirlin
\cite{Mirlin}.

In this Letter, we would like study another interesting statistics, namely 
the statistics of the Riccati variable of wave functions (see, 
e.g., \cite{Luck}).  
The Riccati variable has been studied in one-dimensional disordered 
system, where it is 
related to the reflection coefficient of waves. As we shall see later, 
the Riccati variable can be easily related to the two-point 
correlation function, but it contains more information than the correlation 
function. Furthermore,
the Riccati variable 
gives information about local fluctuation properties of 
wave functions, and the form of its distribution is related to 
properties of the perturbation of the physical models.  
To seek a general universality of this quantity among different
quantum systems, our study will be extended to a wide range of 
random matrix model, disordered and chaotic systems
such as the Random Matrix Model (RMM)\cite{Mehta}, 
Band Random Matrix Model 
(BRMM)\cite{FM94}, the Lipkin model \cite{Lipkin} with chaotic 
behavior in the classical limit, 
a chaotic quantum billiard model\cite{Bunim} and
a 1D tight-binding model in the case of weak disorder\cite{AnTh}.

In order to study the chaotic systems, 
the Riccati variable for the 1D tight-binding model is first extended
to a series types of Riccati variable, 
namely, type I, type II and so on. 
Then, the statistics for type I
Riccati variable is 
found analytically to be of a Lorentzian form for the RMM. Based on 
this fact, we conjecture  that the distribution function of the
Riccati variable of type I is universal for all chaotic and 
disordered 
systems. Our numerical results on several such models support the 
conjecture.
Furthermore, the form of the distribution of the other types 
of Riccati variable gives information 
on properties of chaotic eigenfunctions which can not be supplied 
by, e.g., the statistics of intensity of wave functions. 

Let's first consider a general case in which the Hamiltonians 
can be written in the form 
\begin{equation} 
H= H_0 + \lambda V \label{H}, 
\end{equation} 
where $\lambda $ is a parameter for adjusting the strength 
of the perturbation. 
Eigenvectors of $H_0$ and $H$ 
will be denoted by $|i\rangle$ and $|\alpha \rangle$ in what follows, 
respectively, i.e., 
\begin{equation}
H_0 |i\rangle = E^0_i |i\rangle \label{HE0},\quad 
H |\alpha \rangle = E_{\alpha } |\alpha \rangle .
\label{HE}
\end{equation}
Usually the perturbation $V$ can be written in the form of $V=\sum _t V_t$ 
with $V_t=U_t+U_t^{\dag }$ and $U_t$ coupling a basis state 
to one of the other basis states only. 
In this Letter, we choose the strength of 
$V_t$ comparable with each other. 
The component of an eigenstate $|\alpha \rangle$ on a basis state 
$|i\rangle$ will be denoted by  
$C_{\alpha i} (= \langle i|\alpha \rangle$). 
For simplicity, we consider the case that 
both $H$ and $H_0$ are time-reversal invariant only, 
so that $C_{\alpha i}$ are real.  

In the 1D tight-binding model
the so-called Riccati variable 
is defined as the ratio of components of eigenfunctions of 
the nearest sites, where the off-diagonal 
elements of the Hamiltonian are non-zero for the nearest sits only. 
A natural extension of the concept to the general case of $H$ in 
Eq.~(\ref{H}) is to define the Riccati variable as 
$p_{\alpha , (i,j)}= C_{\alpha i} / C_{\alpha j}$ for $|i\rangle $
and $|j\rangle $ coupled by a $V_t$ ($\langle i|V_t|j\rangle \ne 0$).
Note that the label $j$ is determined by the label $i$ due to 
$V_t$. 
The distribution of $p_{\alpha , (i,j)}$ 
for all possible $i$ of an eigenfunction, or of eigenfunctions with 
eigenenergies in some energy range,   
will be denoted by $f_E(p)$ (the subscript $E$ will be omitted 
for brevity in what follows). 

Furthermore, the ratio $p_{\alpha , (i,j)}$ for $|i\rangle $ and 
$|j\rangle $ not coupled by $V$ is also of interest here. When 
$\langle i|V|j\rangle =0$ but $\langle i|V^2|j\rangle \ne 0$, 
the ratio $p_{\alpha , (i,j)}$ will be called type II 
Riccati variable. The Riccati variable mentioned above 
for $|i \rangle $ and $|j\rangle $ coupled by $V$, is 
type I. 
For $|i \rangle $ and $|j \rangle $ satisfying 
$\langle i|V^n|j\rangle \ne 0$ but $\langle i|V^{n-1}|j\rangle = 0$,
the ratio $p_{\alpha , (i,j)}$ will be called type $n$  
Riccati variable. The distribution of type $n$  
Riccati variable is also denoted by $f(p)$. 
Expressing $C_{\alpha i}C_{\alpha j} $ as $p_{\alpha ,
(i,j)} C^2_{\alpha j} $, one can see that for chaotic eigenfunctions,
whose components and ratios of components can be regarded as statistically 
independent, the average value of $p_{\alpha ,(i,j)}$ 
is directly related to the two-point correlation function of a fixed 
distance by the relation 
$\langle p_{\alpha ,(i,j)} \rangle \langle C^2_{\alpha j} \rangle 
= \langle C_{\alpha i}C_{\alpha j} \rangle $
where $\langle C^2_{\alpha j} \rangle =1/N$. 

{\it Random Matrix Model.} -- We start our discussion with the
RMM\cite{Mehta}. Choosing $\lambda =1 $ in Eq. (\ref{H}) and taking the
matrix elements of $H_0$ and $V$ in the $H_0$ representation to be
real random numbers with Gaussian distribution
($\langle H^2_{ik}\rangle = 1$), one has the Gaussian orthogonal 
ensemble (GOE). The joint distribution of two arbitrary components 
 $C_i$ and $C_j$ for an arbitrary eigenfunction is 
(subscript $\alpha $ omitted for brevity) 
\begin{equation} P(C_i,C_j) = \frac{ \Gamma ((N+1)/2)}{\pi 
\Gamma ((N-1)/2) } (1-C^2_i - C^2_j)^{(N-3)/2} 
\label{PCC} \end{equation} 
(see ref. \cite{Haake}).
Then, the distribution $f(p)$ can be readily obtained 
\begin{equation} 
f_{GOE}(p) = 
\int dC_i dC_j \delta (p - \frac{C_i}{C_j})P(C_i ,C_j) 
= \frac{ 1/ \pi }{p^2 +1}, 
\label{fp1} 
\end{equation}
which is a Lorentzian 
distribution centered at $p=0$. Notice that the ratio of 
any two components of eigenfunctions of the GOE is a type I Riccati 
variable.

It is then natural to conjecture that 
the distribution of type I Riccati variable for 
chaotic eigenfunctions also has a Lorentzian form.  
As will be shown below  numerically
for other models, the distribution of 
the Riccati variable of type I for chaotic eigenfunctions
is in fact of a more general Lorentzian form, denoted by $f_L(p)$ 
in order to be distinguished from $f_{GOE}(p)$,
which may center at $p \ne 0$, 
\begin{equation} 
f_L(p)= \frac{a/\pi}{(p-s)^2+ a^2},
\label{lorentz} 
\end{equation} 
where $a=\sqrt{1-s^2}$ and $s =\langle p\rangle $ which is proportional 
to the value of 
two point correlation function for a fixed $|i-j|$.  In fact, $f_L(p)$ 
can be obtained under a condition more general than the RMM. 
Consider a distribution 
$f(p)$ for $p$  and the distribution $F(q)$ for $q=1/p$. 
If $F(q)$ has the same form as $f(p) $, i.e.,
$F(q)=f(q)$, and $1/f(p)$  can be expanded as a 
Taylor series 
$1/f(p) = d_0 + d_1p + d_2 p^2 + 
\cdots, $
then $f(p)$ must has the form of $f_L(p)$. 
This can be proved  easily  
by making use of the relation between $F(q)$ and $f(p)$ 
required by $q=1/p$, i.e., 
$F(q)=f(1/q)/q^2.$

{\it Lipkin Model.} -- The first dynamical model employed 
in the Letter 
is a three-orbital schematic shell 
model, called the Lipkin 
model, which is bounded and conservative. 
In this model there are totally $\Omega $ particles 
distributed in three orbitals. 
Both the quantum and the corresponding 
classical dynamics of the model 
have already been known (see, e.g., Refs. \cite{MKZ88,WIC97}). 
The Hamiltonian used in Ref. \cite{WIC97} has the form
of Eq. (\ref{H}) with 
\begin{equation}
H_0=\epsilon_1 K_{11} + \epsilon_2 K_{22},\qquad
\displaystyle V= 
\sum_{t=1}^4 \mu_t V_{t}, 
\label{H0} 
\end{equation}
where
\begin{eqnarray}  
V_{1}= K_{10}K_{10}+K_{01}K_{01},\,\quad V_{2}= 
K_{20}K_{20}+K_{02}K_{02},\nonumber\\
V_{3}= K_{21}K_{20} + K_{02}K_{12},\, V_{4}= K_{12}K_{10}+ 
K_{01}K_{21}. 
\label{V14} 
\end{eqnarray}
Here $\epsilon_i$ and $\mu_t$ are parameters given in 
\cite{WIC97}, for which the system 
is almost chaotic when $\lambda =2$. 
The operators $K_{00}, K_{11}$ and $K_{22}$ are   
particle number operators
of the orbitals 0, 1 and 2, and
$K_{rs}$ for $r \ne s$  are particle raising
and lowering operators, respectively.
The eigenstates of $H_0$ are 
$|mn\rangle = A(m,n) K^m_{10} K^n_{20} |00\rangle,$
where $A(m,n)$ is the normalization coefficient, $m$ and
$n$ are particle numbers of the orbitals 1 and 2, 
respectively. The total particle number
is conserved, and as a result the particle number of 
the orbital 0 is $\Omega -m-n$. In our calculations 
we take $\Omega =40$ and the dimension of the Hilbert 
space is $N=(\Omega +1)(\Omega +2)/2= 861$. 

The basis states $|mn\rangle$ can also be labeled in energy 
order as $|i\rangle$. In $|i\rangle$-representation, the Hamiltonian 
matrix is banded \cite{WIC97}. 
For convenience, we can put the two labels  
together with $m_i$ and $n_i$  
indicating particle numbers of the orbital 1 and 2, respectively.
Eq. (\ref{V14}) 
tells that the perturbation 
couples only four kinds of basis states with 
definite values of $\Delta m=m_i-m_j$ and $\Delta n
=n_i - n_j$, e.g.,  
$\Delta m= \pm 2 ,\Delta n=0$ for $V_1$. 
That is, there are totally four kinds of 
Riccati variable of type one.  

When $\lambda =2$, in the middle energy region the 
nearest-level-spacing distribution of the Lipkin 
model is close to the Wigner distribution and the underlying
classical dynamics is almost chaotic (except a 
few quite small regular islands). 
For this perturbation strength, it has been found that 
the distributions of all the four kinds of Riccati variable 
of type I are quite close to the distribution $f_L(p)$. 
As an example, in Fig. 1(a) 
we present the distribution $f(p)$ (dots, in the logarithm scale) 
for $p=C_{\alpha i}/C_{\alpha j}$ 
with $|i\rangle$ and $|j\rangle$ coupled by $V_{1}$. 
In order to have a good statistics 
we have diagonalized 
51 Hamiltonian matrices with $|\lambda -2| \le 0.025$
and then put data obtained for eigenfunctions of  
$\alpha =430 - 450$ together. 
The solid line in Fig. 1(a) 
is the best fitting curve of the Lorentzian
form $f_L(p)$. The central parts of the $f(p)$ and the 
fitting curve are presented in the inset of Fig. 1(a)
in the ordinary scale. 
We would like to mention 
that the center of the distribution $f(p)$ 
in Fig. 1(a) is not at $p=0$. 
 
On the other hand, 
for  Riccati variable of type $n$ with 
$n \ge 2$ (also for $\lambda =2$), it has been found that 
the distribution $f(p)$ deviates from the Lorentzian form $f_L(p)$,  
and the larger $n$  the 
larger the deviation is. In Fig. 1(b) an example 
is given for the distribution of
type V Riccati variable
with $m_i - m_j= \pm 7$ and $n_i - n_j =0$, 
where deviation of $f(p)$ (dots) from the Lorentzian fitting 
curve (solid line) is quite obvious. 
However, when the perturbation strength 
$\lambda $ is increased further, 
the deviation from $f_L(p)$
will become smaller. In fact, for each type of Riccati variable, 
so long as the value of $\lambda $ is large enough, the distribution 
function will become Lorentzian. 
For example, when $\lambda =6$ the form of $f(p)$ for 
type II Riccati variable can be fitted  
as well as the $f(p)$ in Fig. 1(a)  
by the Lorentzian 
distribution $f_L(p)$.

Finally, we would like to mention that 
if the system is not chaotic 
the distribution $f(p)$ for all types of Riccati variable 
deviates from the Lorentzian 
distribution $f_L(p)$. The deviation increases with decreasing
$\lambda $. When $\lambda $ approaches to zero, namely, the system 
becomes near integrable, the distribution $f(p)$ will approach to a 
$\delta $ function. 
 
{\it Band Random Matrix Model.} --
Since the difference 
between the Lipkin model 
and the RMM lies in 
the band structure of the Hamiltonian matrix 
of the Lipkin model in $H_0$ representation, 
we have also studied the distribution $f(p)$ for 
the Band Random Matrix Model (BRMM). 
For this model we take $\lambda =1$ 
in Eq. (\ref{H}) and $H_{ik}$ being 
 real random numbers with Gaussian distribution
($\langle H^2_{ik}\rangle = 1$) for $|i-k| \le b$ and zero otherwise. 
When $b>1$, despite of whether the eigenfunctions are localized or 
extended,  numerically it has 
been found that the distribution $f(p)$ for Riccati variable of type I
($|i-j| \le b$) is very 
close to $f_{GOE}(p)$ (due to the randomness of $H_{ik}$ the center of 
$f(p)$ is at $p=0$);
while for the distribution of Riccati variable of type 
$n$ with $n >1$, 
deviation of the $f(p)$ from $f_{GOE}(p)$ has been found  
which enlarges gradually as $n$ increases. 
An example is given in 
Fig. 1(c) for $i-j= \pm 4b$, i.e. type IV in which the difference 
between the distribution 
$f(p)$ (dots) and the fitting curve of the 
Lorentzian form  (solid line) 
is quite clear. 
When $\lambda $ is increased, similar to the case of the Lipkin 
model, $f(p)$ distributions for Riccati variable of type $n$ with 
$n >1$ will also approach to the Lorentzian form of $f_{GOE}(p)$. 

In order to compare with the result for the Lipkin model 
that centers of $f(p)$ are not necessarily at $p=0$, 
we have tried to add 
a constant to all the non-zero off-diagonal elements 
of the BRMM. Then, it has been found that 
the peak of $f(p)$ could indeed be 
shifted (see Fig. 1(d)) 
to $s(\ne 0)$. The value of $s$ has been found numerically in a good 
agreement with two-point correlation function. 

{\it Chaotic Quantum Billiard.} -- 
Chaotic quantum billiards have been 
widely studied as a prototype of quantum chaos. We take the 
Bunimovich stadium\cite{Bunim} as an example. A chaotic odd-odd 
wave function of a stadium 
with radius $R=1$ and $a=1$ has been calculated by plane wave 
decomposition method\cite{Li}. The 
wave function is shown in the inset of Fig. 2(a). For this wave function 
the statistics of the intensity $|\psi|^2$ is a Porter-Thomas 
distribution. The distribution $f(p)$ for $p=\psi(x,y)/\psi(x,y + \delta 
y)$ with a fixed $\delta y$ larger than the de 
Broglie wavelength is shown in the figure, which is found to 
be a very good Lorentzian.  The best fit by Eq. (\ref{lorentz}) gives 
rise to $s=\langle p\rangle=0.16$. As was pointed out before, $s$ is 
related to the two point correlation function at a fixed $\delta y$. 
According to Berry's conjecture\cite{Berry77}, $\langle 
\psi(x,y)\psi(x,y+\delta y)\rangle = J_0(k\delta y)$, where ${\cal A}$ 
($=1+\pi/4$) is the area of the billiard, and $k\delta y=10\pi$ here. 
Thus, we have $\langle p\rangle = {\cal A}J_0(k\delta y)=0.18$. 
This value is very close to the one obtained above.

{\it 1D Tight Binding Model.} -- Finally, although  statistical 
properties of spectra and distribution of components of eigenfunctions 
for classically chaotic quantum systems 
are quite different from those of the 1D tight binding 
model (Anderson model) in the case of weak disorder, 
it has been found that for the tight-binding model
the distribution of Riccati variable of type I
is also close to the Lorentzian form. 
For this model we take $v_{ik}= \delta _{i,k+1} 
+ \delta _{i,k-1}$ and $E^0_i$ being random numbers
with flat distribution in the region $[-1/2,1/2]$. 
An example is given in Fig. 2(b) for the 
case of $\lambda =1.0$. 
The agreement between 
the $f(p)$ distribution (dots) 
and the fitting curve of the Lorentzian 
form $f_L(p)$ (solid lines)
is quite obvious. 

In conclusion, 
for all different chaotic and disordered models studied in this Letter, 
we have shown that 
the distribution $f(p)$ for type I Riccati variable 
is universal and has the 
form of Lorentzian distribution. 
On the other hand,  the distribution function for   
Riccati variable of type $n$ with $n>1$, 
although approaching to the Lorentzian form when 
perturbation is strong enough, 
shows different features in different stages of perturbation strength
even though the systems are already chaotic. 
Therefore, 
statistics of Riccati variables of different types 
reflects statistical properties of eigenfunctions 
that can not be reflected 
by the statistics of wave function amplitude. 
Furthermore, the properties of Riccati variable may supply 
an alternative measure to probe
information for coupling structure (the Hamiltonian structure) 
for real systems. 

\stars
We are very grateful to G. Casati,  G.-S Tian and P.-Q
Tong for valuable discussions, J. M. Luck for introducing
us his interesting works on Riccati variables. We also thank the referees 
for stimulating comments. This work was supported in parts by
the Hong Kong Research Grant Council (RGC), the Hong Kong
Baptist University Faculty Research Grant (FRG), Natural Science
Foundation of China, and the National Basic Research Project ``Nonlinear
Science'', China.

\begin{figure} 
\caption{(a) The distribution function $f(p)$ (dots) of type I 
Riccati variable
for the Lipkin model in chaotic region. 
(b) Same as (a) but for type V. (c) The $f(p)$ of
type IV Riccati variable for the Band Random Matrix model, i.e. $i-j=20$ 
band width $b=5$. (d) The $f(p)$ (dots) obtained by 
adding a constant to all the non-zero off-diagonal matrix
elements of BRMM. The solid curves in (a)-(d) are best 
fitting curves of the Lorentzian form.  }
\label{fig1}
\end{figure}

\begin{figure} 
\caption{(a) The distribution function $f(p)$ 
(dots) of a chaotic eigenfunction 
($k =\sqrt{E} = 200.119~670$) in stadium billiard. 
The eigenfunction is shown in the 
inset. (b) The $f(p)$ distribution  (dots)
of the 1D Anderson model in the case of weak 
disorder.  The solid curves in (a) and (b) are  best fitting 
curves of the  Lorentzian form.} 
\end{figure}

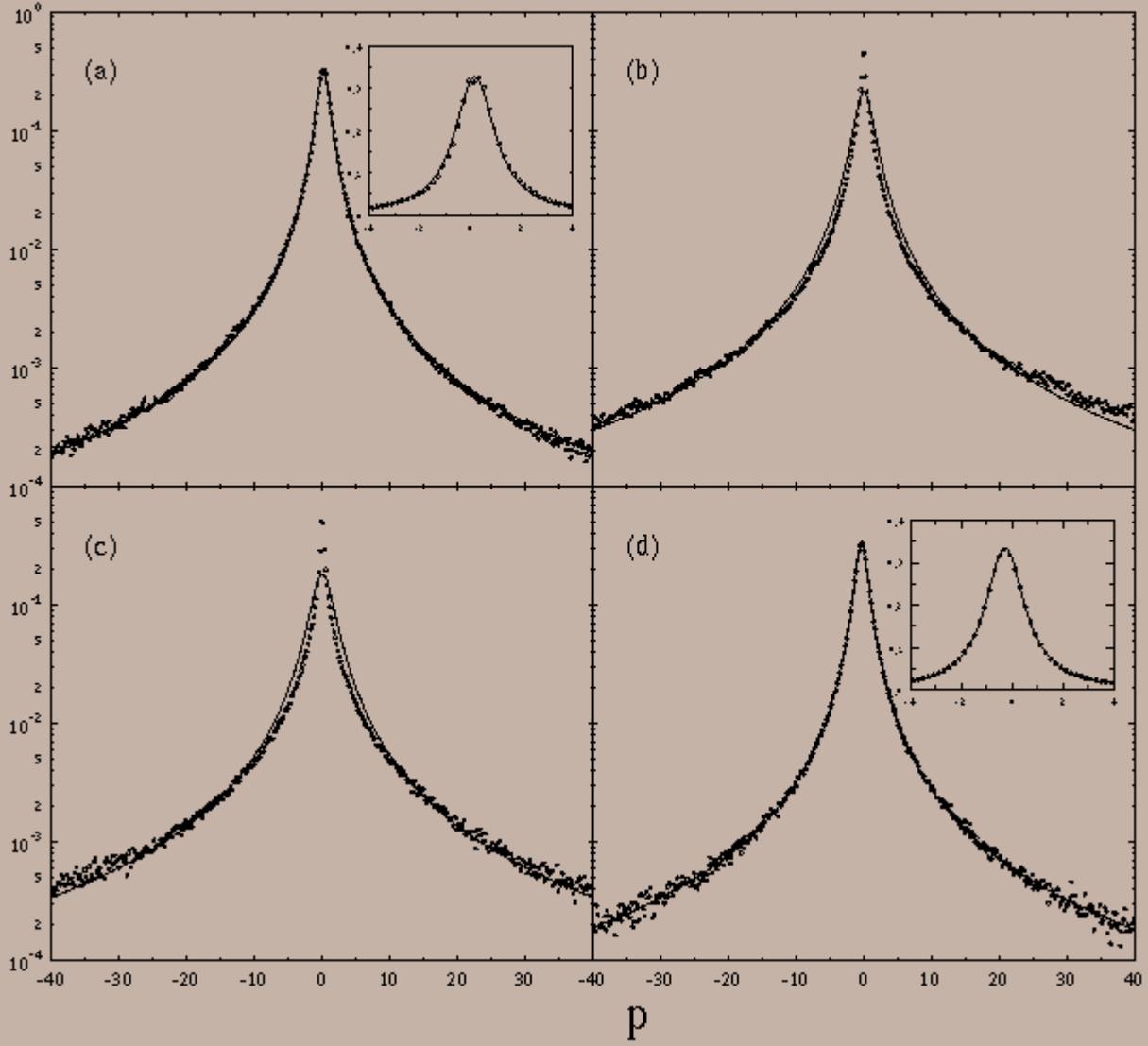
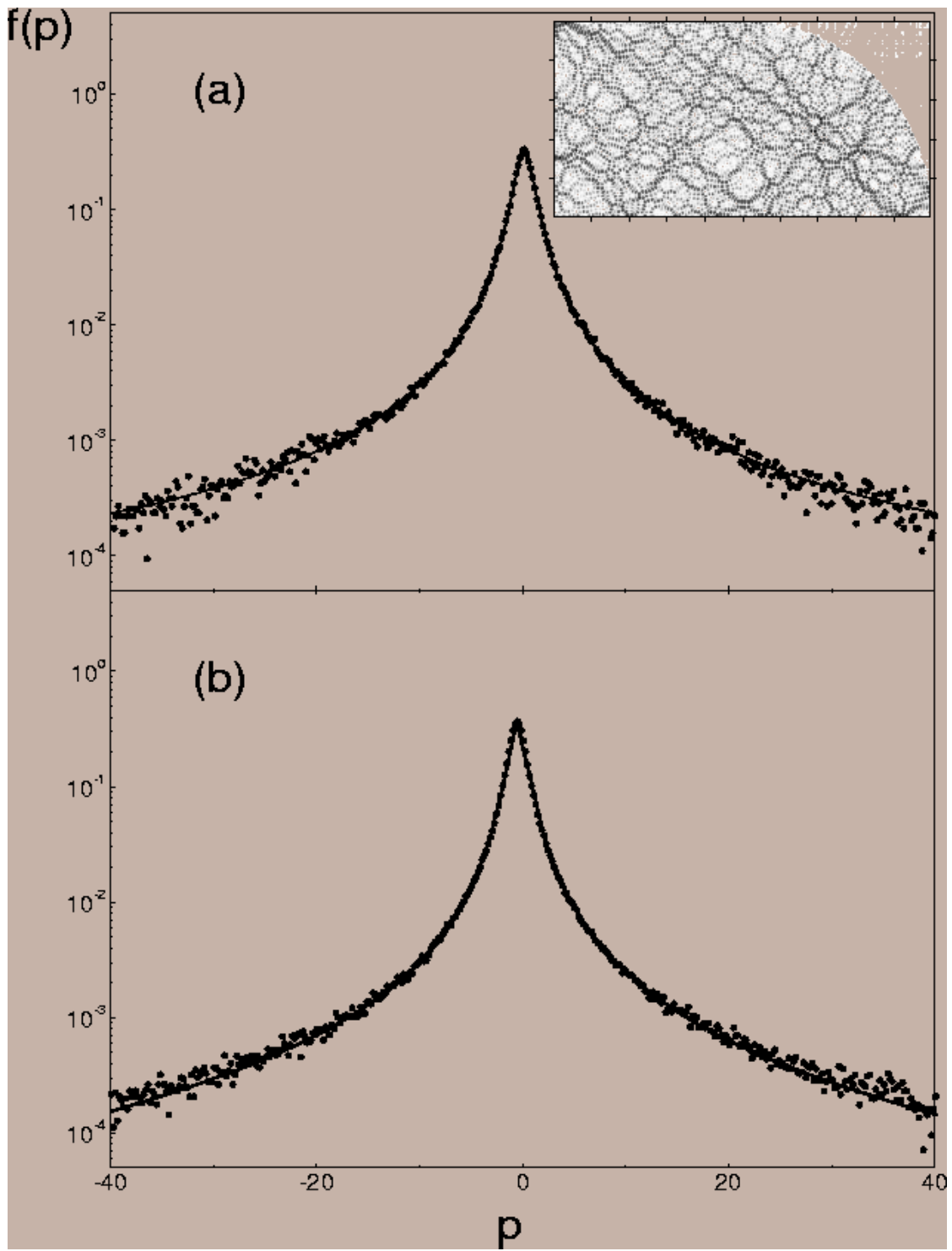
\end{document}